\documentclass[a4paper,twocolumn,fleqn]{article}
\usepackage{amsmath}
\usepackage{amssymb}
\usepackage{txfonts} 
\usepackage{cite}
\usepackage{graphicx}
\usepackage{pfr}
\newcommand\ion[2]{#1\,{\scshape #2}}

\begin{document}
\title{\ion{Fe}{XVIII--XXIV} K$\beta$ Inner-shell Absorption Lines in the 
X-ray Spectra of Neutron Star and Black Hole Binaries with XRISM}
\author{
Masahiro~TSUJIMOTO\sup{1,2}, 
Daiki~MIURA\sup{1,2},
Hiroya~YAMAGUCHI\sup{1,2},
Ehud~BEHAR\sup{3},
Chris~DONE\sup{4},
Mar\'{i}a~D\'{I}AS~TRIGO\sup{5},
Chamani~M.~GUNASEKERA\sup{6},
Peter~A.~M.~VAN~HOOF\sup{7},
Stefano~BIANCHI\sup{8},
Maryam~DEHGHANIAN\sup{9},
Gary~J.~FERLAND\sup{9}
}

\affiliation{
\sup{1}JAXA, ISAS, Sagamihara, Kanagawa, Japan.
\sup{2}The University of Tokyo, Bunkyo-ku, Tokyo, Japan.
\sup{3}Department of Physics, Technion, Haifa, Israel.
\sup{4}Department of Physics, Durham University, Durham, UK.
\sup{5}ESO, Garching bei M\"{u}nchen, Germany.
\sup{6}STSci, Baltimore, MD, USA.
\sup{7}Royal Observatory of Belgium, Brussels, Belgium.
\sup{8}Universit\`a degli Studi Roma Tre, Rome, Lazio, Italy.
\sup{9}University of Kentucky, Lexington, KY, USA.
}
\date{(Received / Accepted )}
\email{tsujimoto.masahiro@jaxa.jp}

\begin{abstract}
 The advent of the X-ray microcalorimeter spectrometer \textit{Resolve} onboard the
 XRISM space telescope opened a new era for high-resolution X-ray spectroscopy of
 astrophysical plasmas. Many spectral features were newly detected, including the
 K$\alpha$ and K$\beta$ inner-shell transition lines of mildly ionized (F- to Li-like)
 Fe at 6--8~keV in the spectra of X-ray binaries and active galactic nuclei. The widely
 used atomic databases contain information on the K$\alpha$ but not K$\beta$ lines of
 these ions. We conducted the atomic structure calculation using \texttt{FAC} to derive
 the Fe K$\alpha$ and K$\beta$ lines and verified the result against ground experiments
 and other calculations of the Fe K$\alpha$ lines. We then implemented the Fe K$\beta$
 lines in a radiative transfer code (\texttt{cloudy}) and compared the synthesized
 and observed spectra with XRISM. A reasonably good agreement was obtained between the
 observation and the \textit{ab initio} calculations. This exemplifies the need to expand
 the atomic databases to interpret astrophysical spectra.
\end{abstract}

\DOI{10.1585/pfr.4.000}

\maketitle

\section{Introduction}\label{s1}
The advent of the \textit{Resolve} instrument\cite{ishisaki2022,kelley2025} onboard the
X-Ray Imaging and Spectroscopy Mission (XRISM)\cite{tashiro2025} opened a new era of
high-resolution (4.5~eV FWHM at 5.9~keV\cite{porter2025}) X-ray spectroscopy in
astrophysical plasmas. It is based on non-dispersive X-ray
microcalorimetry\cite {McCammon1984} as opposed to the previous high-resolution X-ray
spectrometers based on dispersive
technologies\cite{canizares2005,brinkman2000,denherder2001} and excels in energy
resolution, throughput, low background, timing accuracy, and bandpass at $\gtrsim 2$~keV.

\begin{figure}[ht!]
 \includegraphics[width=1.0\columnwidth]{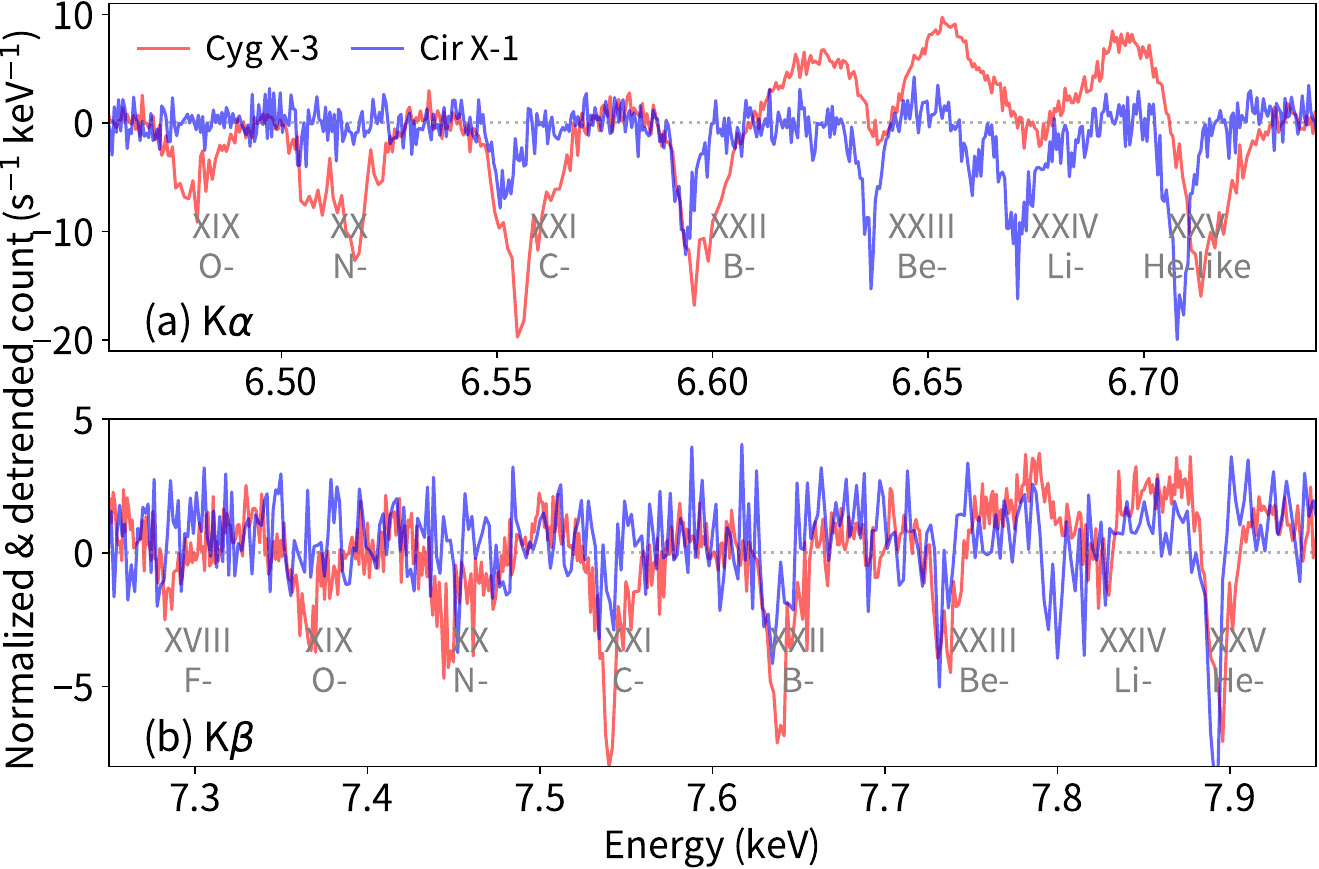}
 \caption{XRISM X-ray spectra (continuum-subtracted and normalized) of Cyg X-3
 \cite{collaboration2024} and Cir X-1\cite{tsujimoto2025} for the (a) Fe K$\alpha$ and
 (b) K$\beta$ features of different charge states.}
 \label{f01}
\end{figure}

Many spectral features were newly detected. One of them is the series of Fe K$\alpha$ ($n=2
\rightarrow 1$) and Fe K$\beta$ ($n=3 \rightarrow 1$) inner-shell excitation absorption
features by mildly-ionized (F- to Li-like) Fe, respectively, in the 6.4--7.0 and
7.3--7.9~keV band (Fig.~\ref{f01}). These lines are observed in active galactic
nuclei and X-ray binaries, which accrete matter into the gravitational potential of the
compact objects (black holes and neutron stars; BH and NS). The released energy is
dissipated into heat and is cooled radiatively. The intense radiation ionizes and
accelerates matter around them, forming outflows of photo-ionized plasmas. This is one
of the common mechanisms for the circulation of matter in the Universe. The Fe K
absorption lines are useful probes of such outflows because of the large cosmic abundance
of Fe and the high enough energy of the lines to penetrate the interstellar
extinction.

What we need to interpret the observed spectra are (i) the atomic data of the
transitions and (ii) the radiative transfer (RT) calculation to synthesize X-ray spectra
under non-local thermodynamic equilibrium (NLTE) conditions. The atomic databases such
as \texttt{AtomDB}\cite{smith2001,foster2017}, \texttt{chianti}\cite{delzanna2021},
\texttt{OpenADAS}\cite{summers2004}, \texttt{xstardb}\cite{mendoza2021}, and the NIST
atomic database\cite{kramida1999}) and the RT codes such as \texttt{xstar}\cite{ko1994},
\texttt{cloudy}\cite{ferland2017}, and \texttt{spex}\cite{kaastra1996,mehdipour2016} are
widely used for astrophysical plasmas. We can constrain the physical properties of
plasmas by comparing observed and synthesized spectra. This approach is generally
established and applies to many lines, but not the K$\beta$ lines of mildly ionized Fe
due to the lack of relevant information in popular atomic databases. Though they were
calculated earlier\cite{behar2022}, only \texttt{chianti} has the Fe K$\beta$ transition
entries for Fe XXIV now. Considering the ubiquity and utility of the K$\beta$ lines
observed with a high signal-to-noise ratio, this needs to be addressed urgently.

This study is one of such attempts. Our approach is to conduct the \textit{ab initio}
atomic structure calculation using a general-purpose code (\S~\ref{s2-1}). We verify the
result against ground measurements and other calculations of the Fe K$\alpha$ lines and
extrapolate them to the K$\beta$ lines (\S~\ref{s2-2}). We run the \textit{ab initio} RT
calculation by adding the Fe K$\beta$ data thus obtained (\S~\ref{s3}) and compare the
synthesized and observed spectra with XRISM (\S~\ref{s4}).

\section{Atomic Structure Calculation}\label{s2}
\subsection{Calculation}\label{s2-1}
We used the \texttt{FAC} (Flexible Atomic Code) version 1.1.5\cite{gu2008} code for the
atomic structure calculation. The code provides a general-purpose, fully relativistic
numerical solver of the Dirac equation with a central field potential. In general, the
$jj$ coupling scheme is used in the relativistic structure calculations, while the LS
coupling scheme is used in the atomic databases. The conversion from $jj$ to LS
coupling was performed using a routine in another structure calculation code
\texttt{GRASP}\cite{fischer2019}.

\begin{figure}[ht!]
 \begin{center}
  \includegraphics[width=0.85\columnwidth]{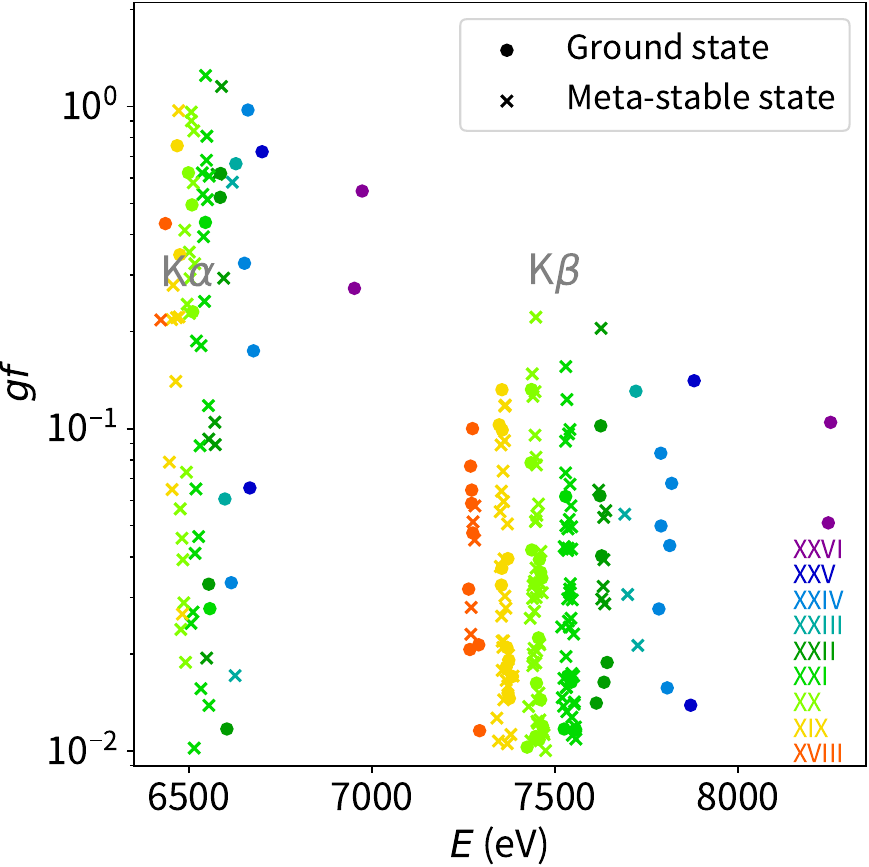}
 \end{center}
 \caption{Scatter plot of the line energy ($E$) and the weighted oscillator strengths
 ($gf$) of the Fe K$\alpha$ and K$\beta$ transitions calculated with
 \texttt{FAC}. Different symbols are used for the transitions from the ground or a
 metastable state. Different colors are used for different ionization stages.}
 \label{f04}
\end{figure}

We calculated the following inner-shell K$\alpha$ and K$\beta$ transitions for the Li-
to F-like Fe:
\begin{description}
 \item[Li-like] 2s or 2p $\rightarrow$ 1s\,2s$^i$\,2p$^{2-i}$ or 1s\,2s$^i$\,2p$^{1-i}$\,$nl$
 \item[Be-like] 2s$^2$ or 2s2p $\rightarrow$ 1s\,2s$^i$\,2p$^{3-i}$ or 1s\,2s$^i$\,2p$^{2-i}$\,$nl$
 \item[B-like] 2s$^2$2p or 2s2p$^2$ $\rightarrow$ 1s\,2s$^i$\,2p$^{4-i}$ or 1s\,2s$^i$\,2p$^{3-i}$\,$nl$
 \item[C-like] 2s$^2$2p$^2$ or 2s2p$^3$ $\rightarrow$ 1s\,2s$^i$\,2p$^{5-i}$ or 1s\,2s$^i$\,2p$^{4-i}$\,$nl$
 \item[N-like] 2s$^2$2p$^3$ or 2s2p$^4$ $\rightarrow$ 1s\,2s$^i$\,2p$^{6-i}$ or 1s\,2s$^i$\,2p$^{5-i}$\,$nl$
 \item[O-like] 2s$^2$2p$^4$ or 2s2p$^5$ $\rightarrow$ 1s\,2s$^i$\,2p$^{7-i}$ or 1s\,2s$^i$\,2p$^{6-i}$\,$nl$
 \item[F-like] 2s$^2$2p$^5$ or 2s2p$^6$ $\rightarrow$ 1s\,2s$^i$\,2p$^{8-i}$ or 1s\,2s$^i$\,2p$^{7-i}$\,$nl$
\end{description}
where $n \in \{3, 4\}$, $i \in \{0, 1, 2\}$, $l \in \{\mathrm{s}, \mathrm{p},
\mathrm{d}\}$ ($n=3$) or $\{\mathrm{s}, \mathrm{p}, \mathrm{d}, \mathrm{f} \}$
($n=4$). Among them, we selected single-excitation and electric-dipole transitions from
the ground state (GS) or a metastable state (MS) of the same quantum number $n$ with the
GS.
The initial states are the following.
\begin{description}
 \item[Li-like] 2s\,$^2S_{1/2}$
 \item[Be-like] 2s$^2$\,$^1S_0$, 2s2p\, $^3P_0$ (43), $^3P_2$ (58)
 \item[B-like] 2s$^2$2p\,$^2P_{1/2}$, $^2P_{3/2}$ (15)
 \item[C-like] 2s$^2$2p$^{2}$\,$^3P_0$, $^3P_1$ (9), $^3P_2$ (15), $^1D_2$ (30), $^1S_0$ (46)
 \item[N-like] 2s$^2$2p$^{3}$\,$^4S_{3/2}$, $^2D_{3/2}$ (18), $^2D_{5/2}$ (23), $^2P_{1/2}$ (33), $^2P_{3/2}$ (41)
 \item[O-like] 2s$^2$2p$^{4}$\,$^{3}P_2$, $^3P_{0}$ (9), $^3P_{1}$ (11), $^1D_2$ (22), $^1S_{0}$ (41)
 \item[F-like] 2s$^2$2p$^{5}$\,$^{2}P_{3/2}$, $^{2}P_{1/2}$ (13)
\end{description}
The excitation energy in eV is given in parentheses for the transitions from MS, all of which
are low enough to be collisionally excited from the predominant GS in the
plasma with a temperature of $\sim$1~MK.
We calculated the energy $E$ (eV), the $A$ coefficient (s$^{-1}$), and the weighted
oscillator strength ($gf$) of these transitions (Fig.~\ref{f04}). The total number of
strong ($gf>10^{-2}$) K$\alpha/$K$\beta$ transitions are 4/6 (Li), 4/4 (Be),
11/14 (B), 23/47 (C), 20/55 (N), 11/40 (O), and 2/14 (F-like).

\subsection{Verification}\label{s2-2}
\begin{figure}[ht!]
 \begin{center}
  \includegraphics[width=0.85\columnwidth]{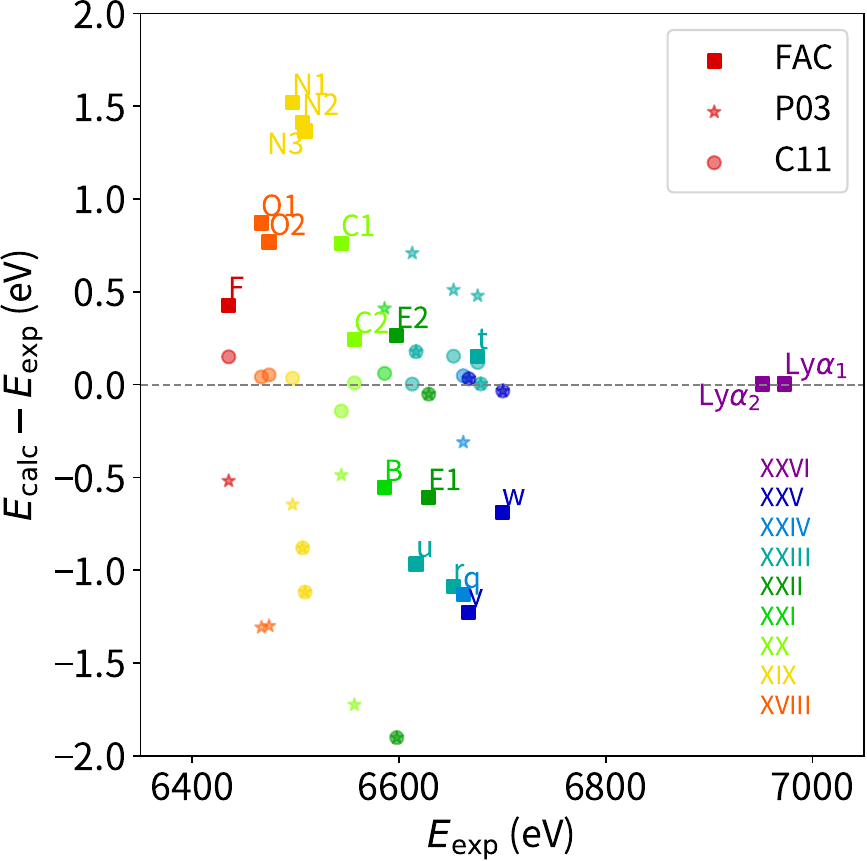}
 \end{center}
 \caption{Scatter plot of the Fe K$\alpha$ line energy of the \texttt{FAC} calculation
 (this work), Palmeri et al. (2003; P03)\cite{palmeri2003a}, and \texttt{chianti}
 version 11.0 \cite{delzanna2021} (C11) against a ground
 experiment\cite{rudolph2013}. The line labels follow \cite{rudolph2013}.}
 \label{f03}
\end{figure}

The energies of the Fe K$\alpha$ inner-shell transitions are well measured in ground
experiments\cite{yerokhin2018,rudolph2013,hell2025} and calculated in
theories\cite{palmeri2003a,palmeri2003b,bautista2003,bautista2004,kallman2004}. We
compared our results with some of them in Fig.~\ref{f03}. The deviation of the
\texttt{FAC} calculation from a ground experiment is within $\pm$1.5~eV
(67~km~s$^{-1}$), which is comparable to another calculation result\cite{palmeri2003a}
and typical uncertainty in energy determination with \textit{Resolve} for mildly-ionized
Fe features\cite{tsujimoto2025}. The mean deviation is 0~eV. We therefore do not apply
any correction to match the \texttt{FAC} results, including the Fe K$\beta$ transitions.

\begin{figure}[ht!]
 \begin{center}
 \includegraphics[width=0.85\columnwidth]{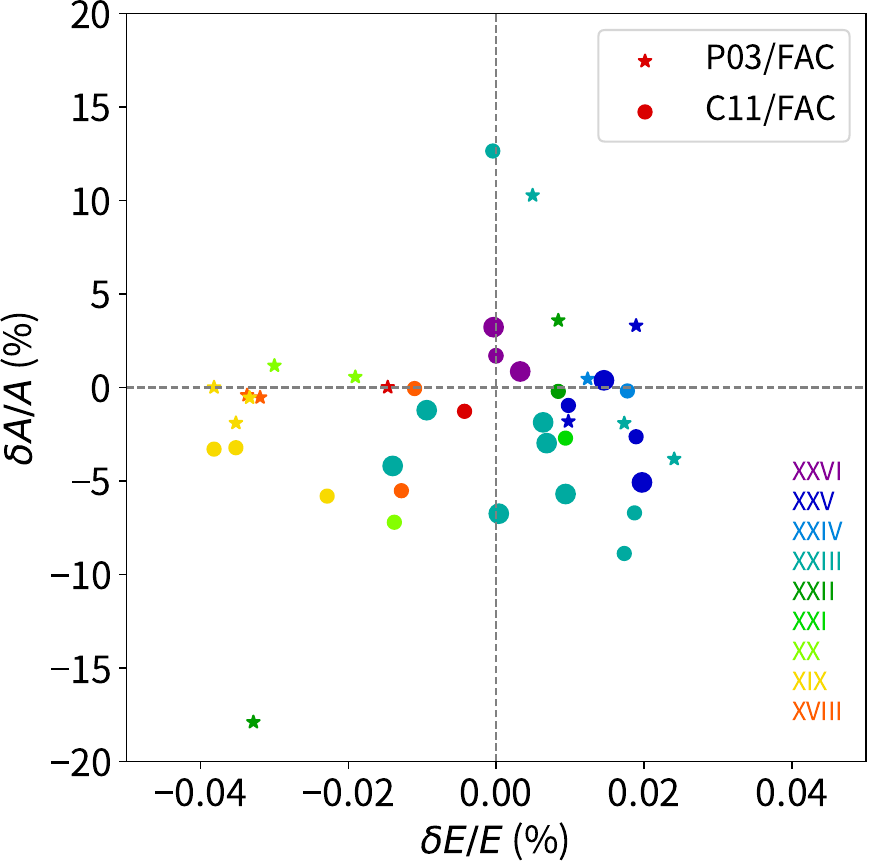}
 \end{center}
 \caption{Scatter plot of relative $E$ and the $A$ value of the Fe K$\alpha$ and
 K$\beta$ (respectively in smaller and larger symbols) between Palmeri et al. (2003;
 P03)\cite{palmeri2003a} and \texttt{chianti} version 11 \cite{delzanna2021} (C11)
 against the \texttt{FAC} calculation (this work).}
 \label{f02}
\end{figure}
We also compared the $E$ and $A$ values of the \texttt{FAC} calculation against another
calculation\cite{palmeri2003a} and the \texttt{chianti} database\cite{delzanna2021} in
Fig.~\ref{f02}. We estimate the systematic uncertainty of $\sim$5\% in $A$ and 0.05\% in
$E$ for the K$\alpha$ (H- to F-like) and K$\beta$ (H- to Li-like) lines. We anticipate
a comparable level of systematics for the other Fe K$\beta$ lines.

\section{Radiative Transfer Calculation}\label{s3}
Plasmas around BH and NS are considered to be in photoionized equilibrium. RT codes
calculate the balances between heating/cooling for the temperature,
ionization/recombination for the charge population, and excitation/deexcitation for the
level population consistently with the external radiation field. All relevant
radiative and collisional interactions need to be
considered. By solving the radiative transfer equation, we obtain the synthesized
spectra of both the direct and reprocessed emission. The direct emission is from the BH
and NS through the plasma, in which absorption features are imprinted upon otherwise
featureless continuum emission. The reprocessed emission is composed of the
electron-scattered emission of the direct emission, and recombination and deexcitation
emission of ions in the plasma.

We used \texttt{cloudy} (c25 release candidate\cite{gunasekera2023}), which is one of
the most widely used RT codes in astrophysics. It adopts the two-stream solver in
one-dimensional setup with the line escape probability approximation without solving the
diffusion across wavelengths. Intensive efforts have been made in recent years to make
it suitable for high-resolution spectroscopy with X-ray microcalorimeters
\cite{camilloni2021,chakraborty2020a,chakraborty2020b,chakraborty2021}, and it is indeed
applied to XRISM data \cite{gunasekera2025,tsujimoto2025,mochizuki2025}. We used
\texttt{chianti} \cite{delzanna2021} for the atomic database among several options with
the method in \cite{gunasekera2022}. Although being most complete, it still lacks the Fe
K$\beta$ inner-shell transition of Be- to F-like Fe. We supplemented them with our
results in \S~\ref{s2}.

\begin{figure}[ht!]
 \begin{center}
 \includegraphics[width=1.0\columnwidth]{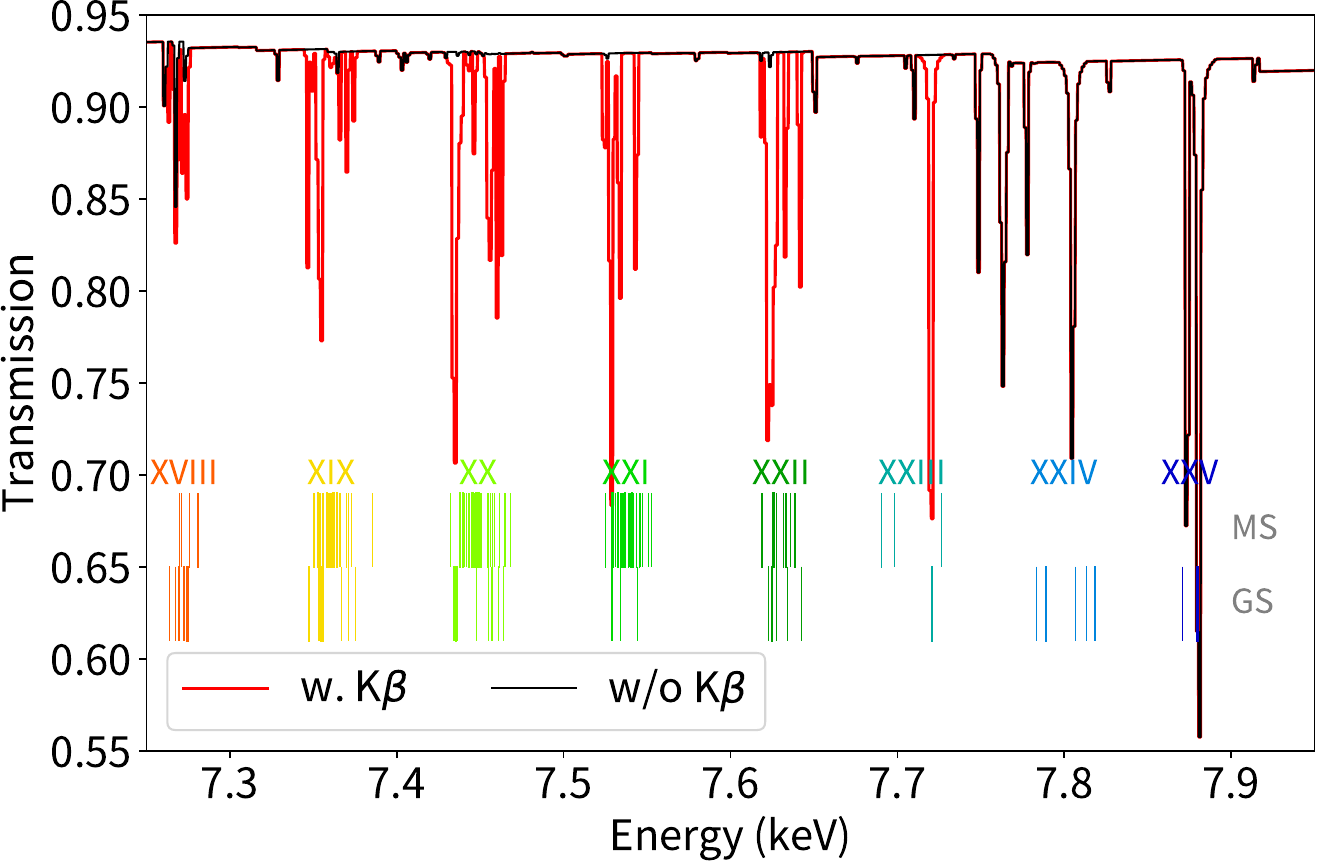}
 \end{center}
 \caption{Transmission through the plasma of $n=10^{12.0}$~cm$^{-3}$,
 $N_{\mathrm{H}}=10^{22.5}$~cm$^{-2}$, and $\xi=10^{2.8}$~erg~cm~s$^{-1}$ calculated
 with \texttt{cloudy} by adding the Be- to F-like Fe K$\beta$ transitions (red) compared
 to the original (black). Line energies of the K$\beta$ transitions are shown by
 vertical lines separately for those from the GS or a MS with a width proportional to
 their $\log{(gf)}$ value.}
 \label{f05}
\end{figure}

We ran \texttt{cloudy} for a grid of three parameters ($n$, $N_{\mathrm{H}}$, and
$\xi$). A plane parallel (slab) geometry was assumed with a constant density of
$n$~cm$^{-3}$ and no turbulence velocity over the slab, and a column density of
$N_{\mathrm{H}}$~cm$^{-2}$ across the slab. The incident radiation is characterized by
the ionization degree $\xi \equiv L_{1-1000}/nr_{\mathrm{in}}^2$~erg~cm~s$^{-1}$, in
which $L_{1-1000}$ is the incident luminosity integrated over 1--1000~Ryd
\cite{tarter1969} and $r_{\mathrm{in}}$ is the distance from the incident source to the
illuminated surface of the slab. The incident radiation has the spectral shape of a
multi-color disk blackbody emission \cite{mitsuda1984} with an innermost temperature of
2~keV, which is typical among X-ray binaries. Fig.~\ref{f05} shows an example of the
synthesized transmitted spectrum.

\section{Comparison to Observations}\label{s4}
We now compare the synthesised and observed spectra. We used the XRISM observations of
Cyg~X-3 (sequence number 300065010) made on 2024 March 24--25 for an integration time of
66.6~ks and Cir X-1 (201036010) on 2025 February 13--15 for 81.5~ks. Cyg X-3 is a binary
of a putative BH and a high-mass star with an orbital period of
4.8~hour\cite{Parsignault1972}, and almost seven whole rotations were covered by
XRISM. Cir X-1 is a binary of a NS and a low-mass star with a period of
16.6~day\cite{Kaluzienski1976}, and a part of a whole rotation (stable
phase\cite{tominaga2023}) was covered. Both are at a distance of $\sim$10~kpc
from the Sun and have an X-ray luminosity of $\sim$10$^{37}$~erg~s$^{-1}$.

Fig.~\ref{f01} shows their \textit{Resolve} spectra in the Fe K$\alpha$ and K$\beta$
bands. Cyg X-3 \cite{collaboration2024} shows both absorption and emission features
stemming from the direct and reprocessed emission, respectively. They have different
Doppler velocity shifts (--530 and 130~km~s$^{-1}$) forming a typical P Cygni
profile. This is a direct consequence of the velocity structure made by the stellar wind
of the high-mass star. The lines are broadened by $\sim$600~km~s$^{-1}$ (12~eV), which
is much larger than the thermal ($\sim$3~eV) and natural ($\sim$0.3~eV) broadening. Cir
X-1, on the other hand, shows only the absorption features. This is because direct
emission overwhelms the reprocessed emission, and thus the emission lines are almost
invisible. The absorption lines are Doppler-shifted and broadened less than Cyg X-3 but
more than the thermal and natural broadening.

\begin{figure}[ht!]
 \includegraphics[width=1.0\columnwidth]{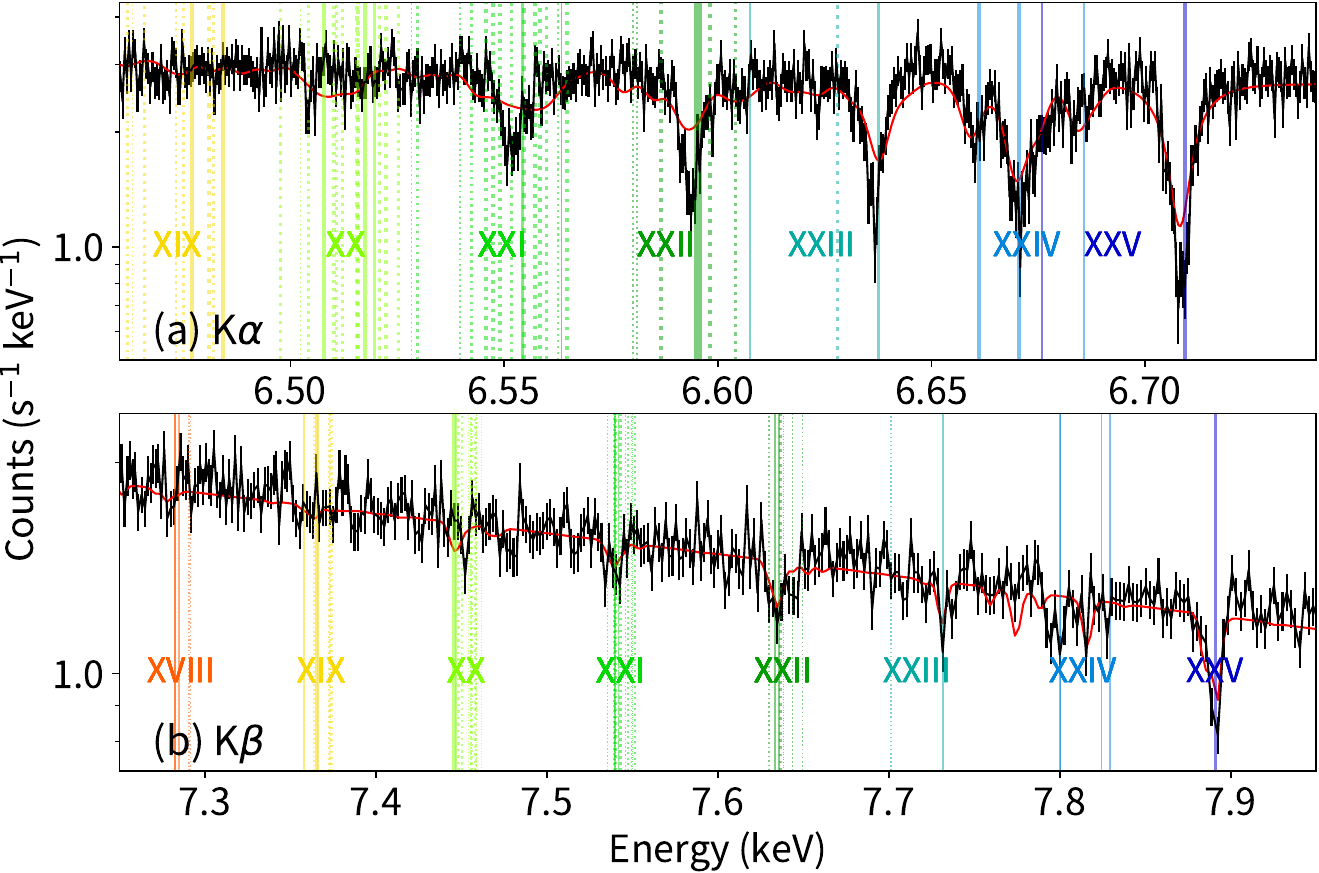}
 \caption{Best-fit RT model (red lines) and the XRISM spectra (black) of Cir X-1. The
 transitions from the GS and MS are given in solid and dotted lines with the energy
 shifted blue-ward by the best-fit Doppler velocity of 425~km~s$^{-1}$.}
 \label{f07}
\end{figure}

We fitted Cir X-1 spectrum with a spectral model consisting of a multi-color disk
blackbody attenuated by the plasma transmission (Fig.~\ref{f05}) and the interstellar
extinction. Lines are shifted for the bulk Doppler motion. A parameter set
($\log{\xi}=3.0$~erg~cm~s$^{-1}$, $\log{n}=14.0$~cm$^{-3}$,
$\log{N_{\mathrm{H}}}=22.5$~cm$^{-2}$) gave a good description overall in both the Fe
K$\alpha$ (Fig.~\ref{f07}a) and K$\beta$ (Fig.~\ref{f07}b) bands. Major observed
features are qualitatively explained, though with some non-negligible deviations. These
deviations provide new constraints on the plasma, such as the $v$, $n$, and
$N_{\mathrm{H}}$ structures as a function of $\xi$ or $r$ once the atomic data are
verified by cross-comparing different calculations and verification with ground
experiments, as was done for Fe K$\alpha$ in \cite{behar2022,palmeri2003a,rudolph2013}.


\section*{Acknowledgements}
Daiji Kato at NIFS reviewed and improved the manuscript.  Gediminas Gaigalas at Vilnius
University and Wenxian Li at CAS provided updates of the \texttt{JJ2LSJ} routine in the
\texttt{GRASP} package. Many participants in the APiP2025 meeting provided
useful comments. This work was supported by the JSPS Core-to-Core Program (grant
number:JPJSCCA20220002) and made use of the JAXA's high-performance computing system
JSS3.


\end{document}